\documentclass[journal]{IEEEtran}

\usepackage{setspace}
\usepackage{hyperref}
\hypersetup{
    colorlinks=true,
    linkcolor=blue,
    filecolor=magenta,      
    urlcolor=blue,
    }
\usepackage{amssymb}
\usepackage{balance}
\usepackage{stmaryrd}
\usepackage{algorithmic}
\usepackage{tikz}
\usetikzlibrary{shapes}
\usepackage{tabularx}
\usepackage{multirow}
\usepackage{arydshln}
\usepackage{lscape}  
\usepackage[justification=raggedright]{caption}
\usepackage{acro}
\acsetup{first-style=short}
\usepackage{array}
\newcolumntype{x}[1]{>{\centering\arraybackslash\hspace{0pt}}p{#1}}
\usepackage{ulem}
\usepackage{color}

\newcounter{Q}
\newcounter{TD}

  \graphicspath{{Figures/}}

  \DeclareGraphicsExtensions{.pdf,.jpeg,.png}

\usepackage{graphicx}

\usepackage{epstopdf}
\usepackage{caption}
\usepackage{subcaption}

\usepackage{amsmath}

\usepackage{authblk}

\begin{document}
\title{Video-Streaming Biomedical Implants using Ultrasonic Waves for Communication}
\author{Gizem Tabak, Jae Won Choi, Rita J. Miller, Michael L. Oelze, Andrew C. Singer\\
University of Illinois at Urbana-Champaign}


\maketitle

\begin{abstract}
The use of wireless implanted medical devices (IMDs) is growing because they facilitate continuous monitoring of patients during normal activities, simplify medical procedures required for data retrieval and reduce the likelihood of infection associated with trailing wires. However, most of the state-of-the-art IMDs are passive and offline devices. One of the key obstacles to an active and online IMD is the infeasibility of real-time, high-quality video broadcast from the IMD. Such broadcast would help develop innovative devices such as a video-streaming capsule endoscopy (CE) pill with therapeutic intervention capabilities.  State-of-the-art IMDs employ radio-frequency electromagnetic waves for information transmission. However, high attenuation of RF-EM waves in tissues and federal restrictions on the transmit power and operable bandwidth lead to fundamental performance constraints for IMDs employing RF links, and prevent achieving high data rates that could accomodate video broadcast. In this work, ultrasonic waves were used for video transmission and broadcast through biological tissues. The proposed proof-of-concept system was tested on a porcine intestine \textit{ex vivo} and a rabbit \textit{in vivo}. It was demonstrated that using a millimeter-sized, implanted biocompatible transducer operating at 1.1-1.2 MHz, it was possible to transmit endoscopic video with high resolution (1280 pixels by 720 pixels) through porcine intestine wrapped with bacon, and to broadcast standard definition (640 pixels by 480 pixels) video near real-time through rabbit abdomen \textit{in vivo}. A media repository that includes experimental demonstrations and media files accompanies this paper \footnote{The accompanying media repository can be found at this link: \url{https://bit.ly/3wuc7tk}}.
\end{abstract}

\IEEEpeerreviewmaketitle

\section{Introduction}
Currently, radio-frequency (RF) electromagnetic waves are the most frequently used method in wireless communication applications such as television, radio, or mobile phone communications. When RF waves travel through the air, they experience little attenuation. Additionally, they can operate at high frequencies, where the available bandwidth is also high. Their capability of operating at high frequencies while experiencing low loss makes RF waves appropriate for long-range, high data rate wireless communication applications through the air. However, there are various drawbacks of using RF waves with wireless IMDs to transmit data through the body. RF waves are highly attenuated in the body and have limited penetration depth \cite{sayrafian2009statistical}. Therefore, higher power levels need to be employed to compensate for the losses in the body due to high attenuation. However, the RF signal power levels that an IMD could deploy are limited for safety reasons, as higher power increases the risk of tissue damage \cite{FCCtelemetry}. There are also federal regulations on the allocation of the RF spectrum use within, and outside the body. According to the Medical Device Radio Communications (MedRadio) guidelines, the allocated operation frequencies for the IMDs are within the range of 401-406 MHz, and the corresponding maximum allowed bandwidth is 300 kHz \cite{FCCradio}. Moreover, MedRadio transceivers are further limited by interference regulations because they must be able to operate in the presence of primary and secondary users in those bands. Such restrictions on the transmit power and operable bandwidth lead to fundamental performance constraints for IMDs employing RF links, and the data rates of the current RF-based IMDs are demonstrated to be limited to 267 kbps \cite{koprowski2015overview}. Considering, for example, the standard definition video requires 1.2 Mbps bitrate, while high definition video streaming starts at 3 Mbps \cite{IBM}, these regulations set a significant barrier against possible wireless IMD applications to include video transmission, which could enable innovative procedures for various IMDs including but not limited to capsule endoscopy devices. 

Capsule endoscopy (CE) pills, i.e., swallowable pills that can capture and record images of the intestines, have revolutionized the medical community’s approach to gastrointestinal (GI) diseases since their introduction two decades ago \cite{iddan2000wireless}. CE provides significant help with diagnosing several conditions, including Crohn’s disease, celiac disease, and small bowel tumor \cite{goenka2014capsule}. It is considered the gold standard to investigate obscure GI bleeding and iron deficiency anemia \cite{mustafa2013small}. 

The state-of-the-art CE systems employ a small capsule pill swallowed by the patient and a data recorder box strapped to the patient. The capsule traverses through the GI tract passively with the help of bowel contractions. While passing, it collects still frames and transmits them to the strapped receiver using radio-frequency (RF) electromagnetic (EM) waves \cite{enns2017clinical}. After the pill passes through the GI system, which could take up to 12 hours, a gastroenterologist investigates the collected images to make a clinical decision \cite{steiger2019ingestible}. CE is considered a safer and less invasive alternative to conventional tethered endoscopy. It does not require for the patient to go under anesthesia. Furthermore, it enables the imaging of small intestines, which are out of reach with tethered endoscopy probes \cite{ciuti2016frontiers}. However, the passive motion of the pill raises the risk of failure to capture portions of the tract \cite{enns2017clinical}. Furthermore, it prevents CE devices from performing therapeutic interventions and acquiring diagnostic tissue samples \cite{ciuti2016frontiers}. Such drawbacks inhibit the broader use of CE in clinical practice \cite{enns2017clinical, steiger2019ingestible, ciuti2016frontiers}. 

\begin{figure*}[t]
\includegraphics[width=\textwidth]{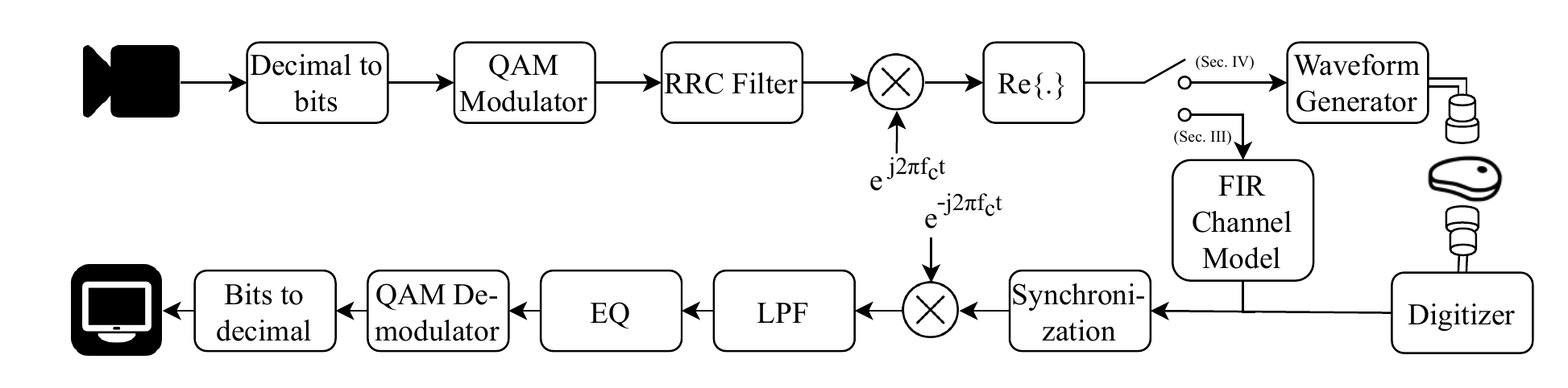}
\caption{Proposed communication system with equalizer (EQ) at the receiver}
\label{fig:framework}
\end{figure*}

For many years, ultrasonic waves have been widely used as an alternative to RF electromagnetic waves in underwater communication applications, where RF waves experience significant losses \cite{riedl2014towards}. The similarities between the ultrasonic communication channels through the body and under water suggest ultrasonic waves as a promising option for wireless communication in biomedical implants. Besides experiencing lower loss and hence propagating deeper in the tissue, ultrasonic waves are desirable for wireless through-body communications for several other reasons: First, because the loss is lower compared to RF electromagnetic waves, the transmission can take place at lower transmit power levels. As a result, the patient experiences lower, if not insignificant, tissue heating. Second, medical applications that utilize ultrasonic waves, such as ultrasonic imaging, have been considered as a safer option compared to applications that utilize electromagnetic waves, such as X-ray imaging, which exposes the patients to significant amounts of ionizing radiation \cite{FDA}. Third, because there are no official regulations on the ultrasonic frequency spectrum, the available bandwidth, and the corresponding potential for high data rates, are significantly higher. For all these reasons, employing ultrasonic waves for through-tissue communications at video-capable data rates offers a safe and efficient alternative to RF communications.

Ultrasonic waves have been used in the literature for wireless in-body and through-body communications. Various studies have demonstrated the feasibility of the ultrasonic communication link through biological tissues. Earlier methods in the literature either achieve lower data rates ($<$1 mbps) that are insufficient for standard video communication \cite{bos2018enabling, wang2017exploiting, chang201727, kondapalli2017multiaccess, santagati2014sonar} or achieve higher data rates with large form factor transducers ($>$1cm) that could not be utilized in a small implantable device \cite{demirors2016high, singer2016mbps}, or the communication link is established through phantoms instead of real biological tissues \cite{kondapalli2017multiaccess, demirors2016high}. More recently, ultrasonic waves have been proposed for a variety of biomedical applications including biomedical imaging, data transmission and drug delivery. In \cite{qiu2020ultrasound}, the authors built a capsule device for ultrasound echo imaging, which enabled imaging within the GI wall as opposed to the surface area typically imaged with the optical CE devices. In that work, the ultrasound data were stored in the pill and retrieved after the procedure. In \cite{stewart2021ultrasound}, the authors built a proof of concept capsule device that demonstrated the feasibility of delivering gastrointestinal therapeutic drugs using ultrasound microbubble agents. In \cite{tabak2021Video}, ultrasonic waves were used to transmit video recordings at video-capable data rates using millimeter-sized transducers through water and \textit{ex vivo} beef liver, \textit{ex vivo} pork chop and \textit{in situ} rabbit abdomen. 

Recent developments encourage the use of ultrasound in CE devices and paves the way for active, online IMDs. However, a real-time or near-real time video broadcasting system that can transmit information from the implanted device inside the body to the receiving device outside of the body is still lacking. In this work, we first establish a video broadcast link through the abdominal wall of an anesthetized rabbit. We show that using the communication system presented in \cite{tabak2021Video}, it was possible to broadcast standard definition video obtained with an endoscopy camera using a millimeter-sized ultrasonic transducer implanted behind the abdominal wall. We then demonstrate video transmission through a saline-filled porcine intestine wrapped with bacon using mm-sized transducers. These experiments reveal the feasibility of ultrasound for broadcasting information near real-time through luminal tissue using mm-sized elements. When combined with a small form factor capsule device with further engineering work, the proof-of-concept system could enable CE devices to have real-time video broadcasting and could be employed in active, online medical implants.

\section{Rabbit Study}
\subsection{Experimental Setup}
\begin{figure}[t]
\includegraphics[width=\columnwidth]{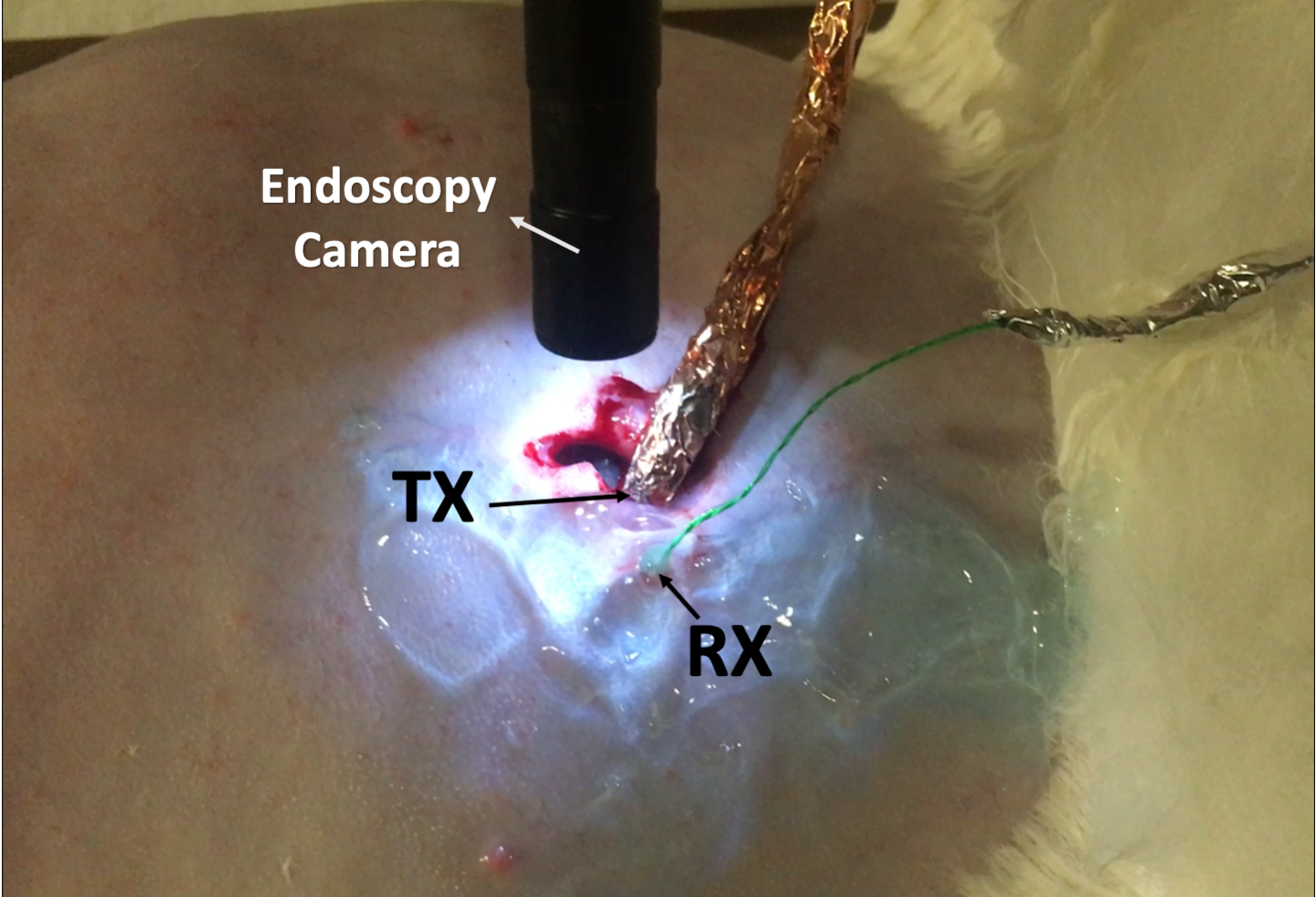}
\caption{Transmitting transducer was placed behind the abdominal wall of the rabbit. Receiving transducer was placed on the shaved abdomen, with gel coupling. The video stream was obtained from an endoscopy camera.}
\label{fig:rabbit-image}
\end{figure}

The video stream is obtained from a capsule endoscopy camera (T01 8.5 mm USB Semi-Rigid Endoscope, Depstech) in .flv format in standard definition (640 pixels x 480 pixels) at 30 frames by second. The bitstream is converted to a QAM transmission signal as described in Sec. \ref{sssec:signal}. The modulated signal was used to drive the transmitting biocompatible sonomicrometry crystals of 2-mm diameter that operate near 1.2 MHz (2-mm Round, Sonometrics, London, CAN). The transmitting transducer was implanted behind the abdominal wall of an anesthethized rabbit. Electrical cables connecting the elements to the equipment were wrapped in copper foil tape for interference shielding. The receiving transducer was placed on the shaved abdomen of the rabbit with gel coupling (Fig. \ref{fig:rabbit-image}). The signal was sent in through-transmission configuration. The distance between the two transducers was 2.2 cm.

An arbitrary waveform generator (PXI-5422, National Instruments, Austin, TX) was used to generate the transmission signal. A differential preamplifier (Model 7000, Krohn-hite Corporation, Brockton, MA) was used to acquire the signal from the receiving transducer configured in differential mode. A digitizer was used (PXI-5124, National Instruments, Austin, TX) to acquire the signal through the differential amplifier. In order to drive the National Instruments equipment and to process the received data in real-time, custom software developed in C\# was used. 

\subsection{Transmission Signal}\label{sssec:signal}
\begin{figure*}[t]
\includegraphics[width=\textwidth]{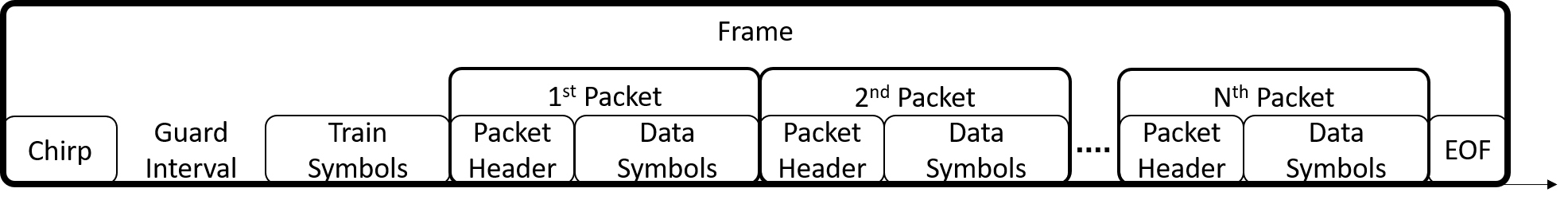}
\caption{Frame structure used for the video streaming. Each frame start with chirp for frame detection and synchronization. Guard interval (silent interval) is placed after the chirp in order to prevent multi-path chirp arrival from interfering with train symbols. A frame consist of $N$ number packets, where $N$ is dependent on the available video data from FFMPEG. After $N$ packets, end-of-frame (EOF) block is inserted in place of a packet header to notify the end of a frame transmission.}
\label{fig:frame_structure}
\end{figure*}

The communication system was presented in detail in \cite{tabak2021Video}. A system diagram is shown in Fig. \ref{fig:framework}. The video stream obtained from the USB camera is converted into .flv format using FFMPEG multimedia framework \cite{tomar2006converting} and pushed into a shared buffer between FFMPEG application and software modem. If there are data present in the FFMPEG shared buffer, the modem streams in as many bytes as possible in multiples of 1024 bytes. The streaming bits are mapped into transmit symbols by the digital modulator. In this work, QAM is used due to its potential for high spectral efficiency. The QAM symbols are upsampled by $L=\frac{f_s}{f_{b}}$, where $f_s$ is the sampling rate, $f_b$ is the symbol rate. The upsampled symbols are then shaped with a root-raised cosine filter $p(t)$, resulting in the data packet
\begin{equation}
    x_D(t)=\sum_{k=0}^{N-1}x_kp(t-kT_b)
\end{equation}
where $T_b=\tfrac{1}{f_b}$ is the symbol period. In order to detect the signal arrival at the receiver, a linear chirp preamble is appended at the beginning of each data frame at the transmitter. The transmission frame, which consists of the linear chirp preamble, guard interval, and data packet, is then mixed with a sinusoidal carrier centered at $f_c$. The passband signal
\begin{equation}
    x(t) = \mathcal{R}e\left\{\sum_{k=0}^{N-1}x_kp(t-kT_b)e^{j2\pi f_c t}\right\}
\end{equation}
is sent to the waveform generator from the software modem through a memory buffer. Figure \ref{fig:frame_structure} summarizes the structure of a transmission frame. In this experiment, the data was sent through the rabbit abdomen with a 16-QAM, 1.2 MHz center frequency signal with 500 kHz symbol rate. 

\subsection{Processing}
At the receiver end, the signal was captured by the digitizer, coarsely aligned and corrected for Doppler effects using the chirp preamble, and the received data packet was decoded using the fractionally-spaced, phase-tracking, sparse DFE \cite{stojanovic1994phase}, \cite{lopez2001dfe}. The equalizer had at most 70 feedforward taps and 200 feedback taps. 4000 known symbols are transmitted per frame for training, and rest were used in the decision-directed mode while updating the equalizer coefficients to track the channel variations. For every 2048 symbols, a packet header symbol is inserted to notify receiver whether the frame transmission ended, or more data blocks are arriving for processing. The coefficients were updated using the recursive least-squares algorithm with learning rate 0.997.

\subsection{Results}
Using the transmission system described above, it was possible to stream SD video near-real time for 25 seconds with approximately 2 seconds of latency. A video recording of the streaming can be found in the \href{https://bit.ly/3wuc7tk}{accompanying repository\footnote{\url{https://bit.ly/3wuc7tk}}} with the file name \texttt{rabbit\_stream.mov}, and a video collage of the setup as well as transmit and receive video files can be found with the file name \texttt{TX-RX-video.mov}, both under the folder \texttt{Rabbit study}.

\begin{figure}[t]
\begin{subfigure}[t]{0.405\columnwidth}
\includegraphics[width=\columnwidth]{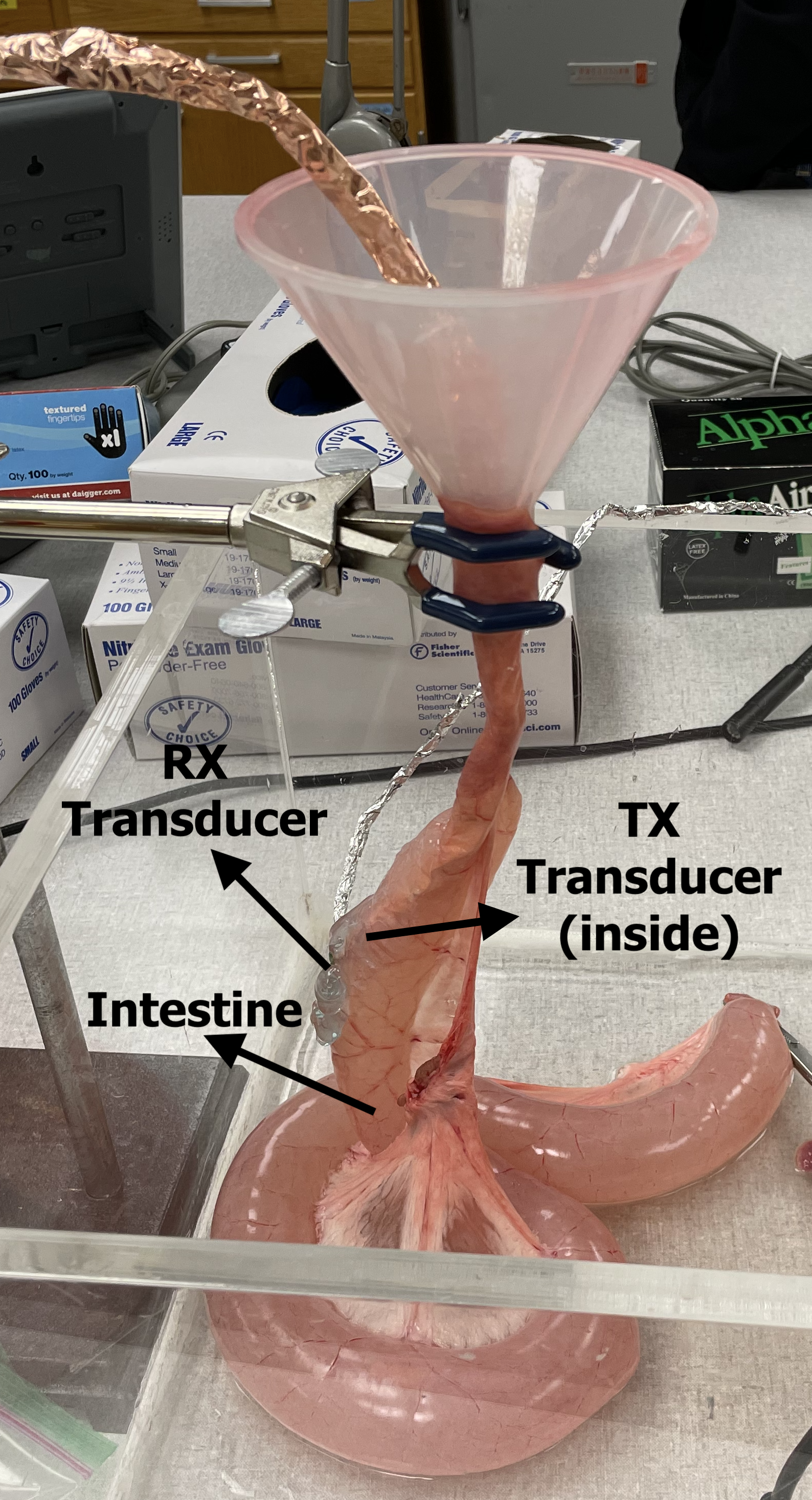}
\caption{Porcine intestine setup}
\end{subfigure}
\begin{subfigure}[t]{0.55\columnwidth}
\includegraphics[width=\columnwidth]{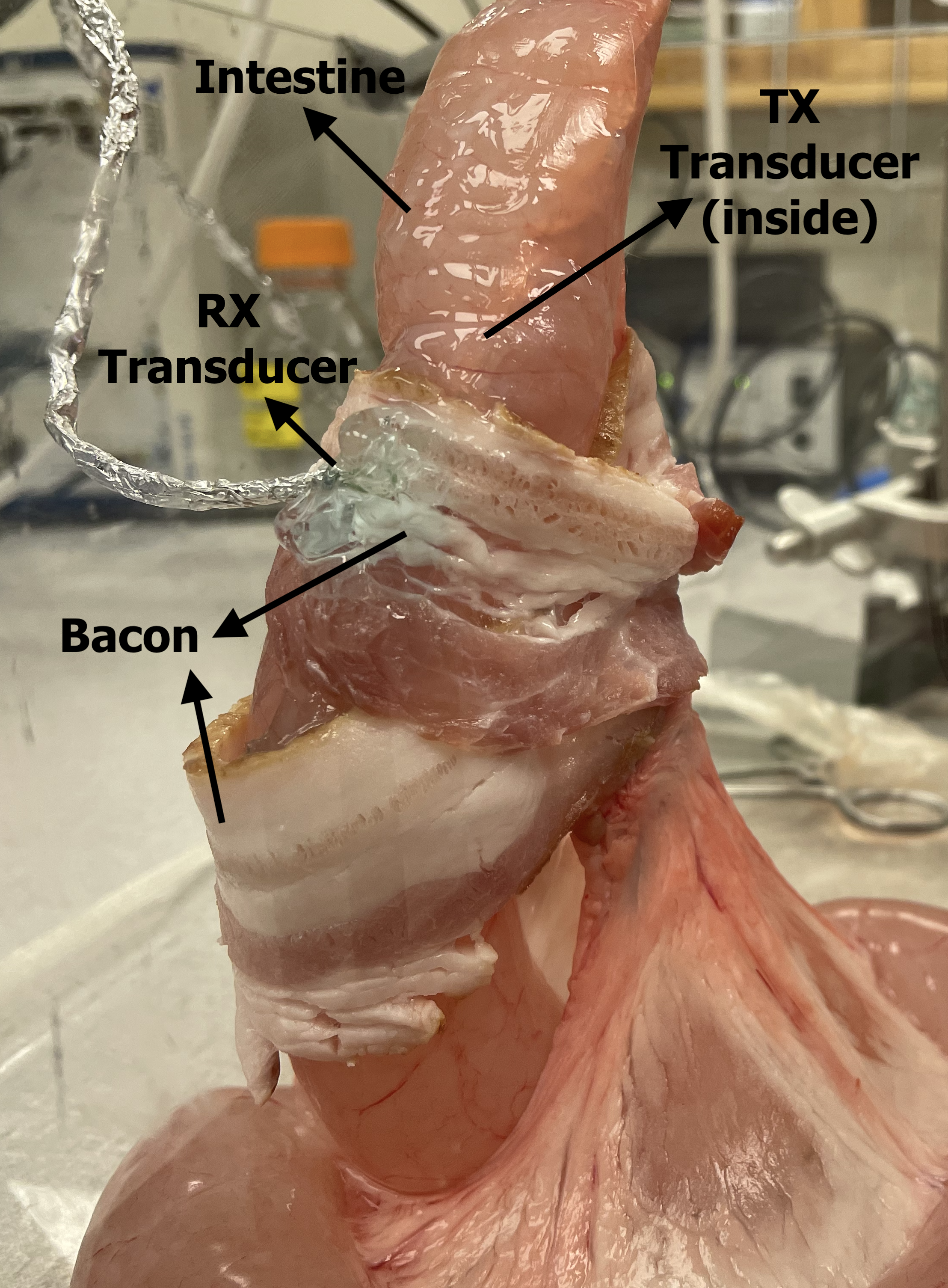}
\caption{Intestine wrapped with bacon}
\end{subfigure}
\caption{Transmitting transducer was placed inside a saline-filled porcine intestine through a funnel. Receiving transducer was placed on the outside surface of the bacon wrapped around the intestine.}
\label{fig:intestine-image}
\end{figure}

\section{Porcine Study}
\subsection{HD Video Transmission}
\subsubsection{Experimental Setup}
In order to test the ultrasonic transmission through a luminal tissue, an approximately 2 feet long \textit{ex vivo} porcine intestine was used. The tissue sample was filled with saline and suspended in an empty tank with a clamp. A slice of bacon was wrapped around the intestine to mimic the surrounding tissues. Transmitting transducer was placed inside the saline-filled intestine through a funnel. Receiving transducer was placed on the outside surface of the bacon wrapped around the intestine. The video used for transmission was obtained using the USB endoscopy camera through the funnel opening and saved in \texttt{mp4} format. The video duration was 1 second, limited by the on-board memory of the acquisition device. In this experiment, a QPSK signal with 1.2 MHz center frequency and 500 kHz symbol rate was used. The benchtop system used for signal transmission and acquisition was the same as the one used in the rabbit study. 

\subsubsection{Results}
\begin{figure}[t]
\includegraphics[width=\columnwidth]{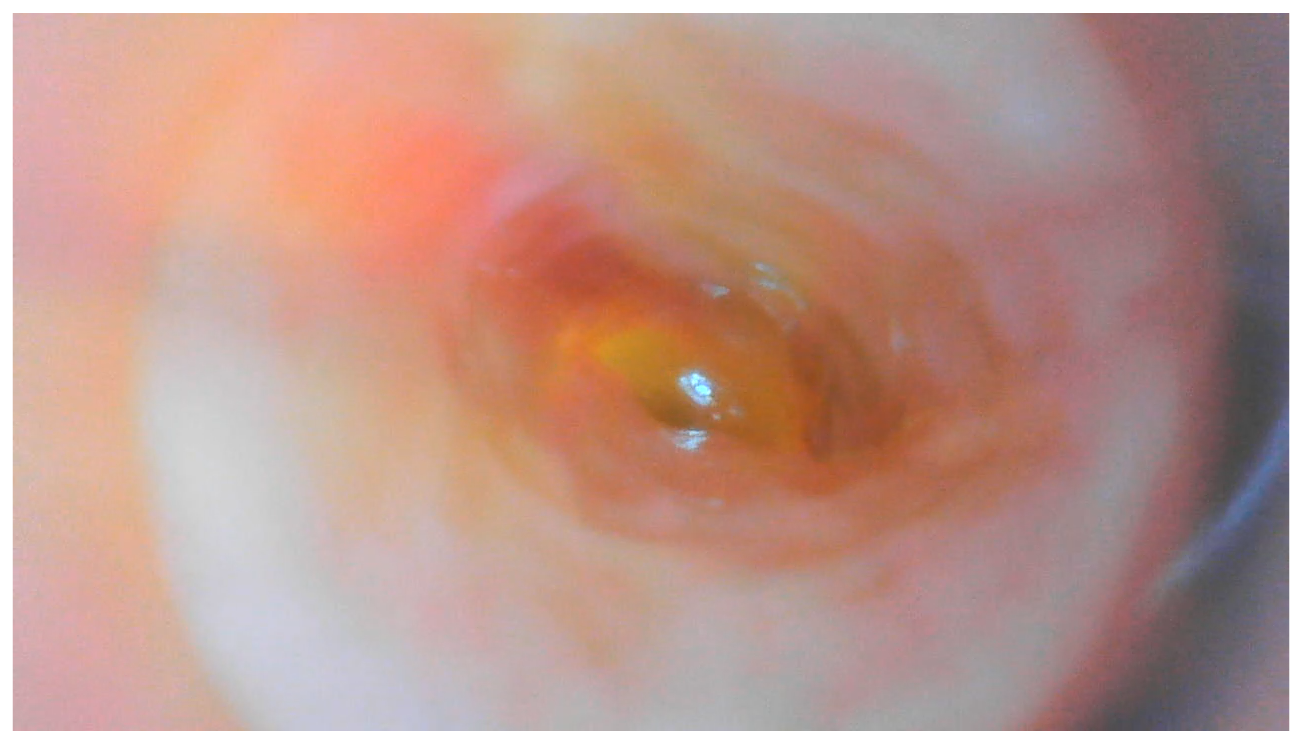}
\caption{A video frame from the received video transmitted with ultrasonic transducers through the intestinal wall wrapped with bacon.}
\label{fig:frame}
\end{figure}

A total of 4,770,000 symbols were transmitted. The DFE had 70 feedforward taps and at most 211 feedback taps. 16\% of the ground truth symbols were used for adaptive filter coefficient update and error correction. For the rest, DFE was used in decision-directed mode (i.e. no ground truth was used). The average bit error rate was 7e-6. The transmitted and received video files can be found in the accompanying repository, with the file names \texttt{TX\_video1.mp4} and \texttt{RX\_video1.mp4}, respectively. An overlay of videos with the experiment setup is displayed in \texttt{TX-RX-video1.mp4} in the folder \texttt{Porcide study}. A representative frame obtained from the received video is displayed in Fig. \ref{fig:frame}.

\subsection{Endoscopy Video Transmission}
\subsubsection{Experimental Setup}
\begin{figure}[t]
\includegraphics[width=\columnwidth]{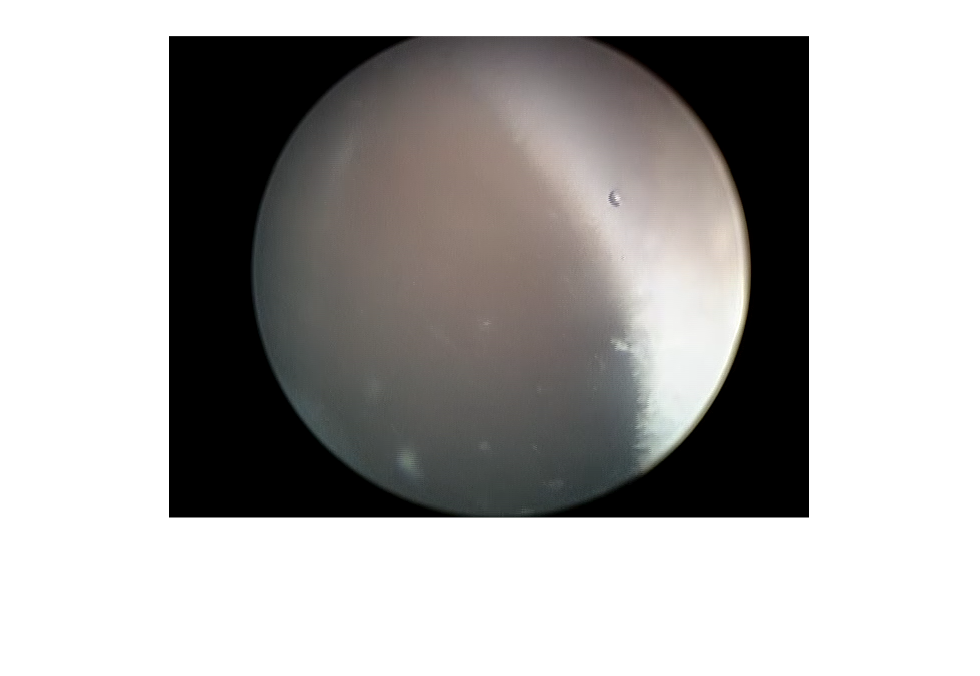}
\caption{A video frame from the medical-grade endoscopy video obtained with the video endoscopy tower system.}
\label{fig:frame2}
\end{figure}

A second experiment was performed to transmit endoscopic video obtained with a video endoscopy tower system (Stryker SDC HD Video Endoscopy Tower System, Stryker Corporation, Kalamazoo, MI). Several endoscopic videos were obtained form inside the suspended intestine filled with saline water. The videos can be found in the \texttt{EndoscopicVideos} folder in the accompanying repository. The video chosen for transmission was 2.5 seconds long and had resolution of 640 pixels by 480 pixels. A frame from the video is shown in Fig. \ref{fig:frame2}. 

The transmitting transducer was inserted inside the intestine through a funnel. The receiving transducer was inserted between two layers of bacon wrapped around the intestine. The raw bytes obtained from the video file was divided into 8 packets and sent through the channel using a 16-QAM signal with 1 MHz symbol rate, centered at 1.13 MHz. Devices used for transmission and receiving were the same with the previous experiment.

\subsubsection{Results}
A total of 1,285,664 symbols that represent the video file were transmitted. In post-processing, the same receiver structure with the DFE was used. Less than 6\% of these symbols were used for training and error correction. The resulting average BER was 1.5e-5 and the data rate was 4 mbps. The additional layer of bacon wrapped around the receiving transducer may have provided the dampening affect for the multipath channel and contributed to higher data rates. The transmitted and received video files can be found in the accompanying repository, with the file names \texttt{TX\_video2.mp4} and \texttt{RX\_video2.mp4}, respectively. An overlay of videos with the experiment setup is displayed in \texttt{TX-RX-video2.mp4} in the folder \texttt{Porcide study}. 

\section{Results and Discussion}
In this work, we presented two proof-of-concept studies that demonstrate the feasibility of using ultrasound to transmit and broadcast video wirelessly from the biomedical implants. First, SD video was broadcasted near real-time through rabbit abdomen \textit{in vivo}. Then, HD video obtained with a commercial endoscopy camera was transmitted through a porcine intestine. Finally, an SD video obtained through a medical-grade video endoscopy tower was transmitted through the porcine intestine. In each case, the videos were sent to a benchtop computer using millimeter-sized, biocompatible transducers. The broadcast in the rabbit experiment was established with a 16-QAM modulated signal with 500 kHz symbol rate, and the HD video transmission through the porcine intestine was established with QPSK-modulated signal with 1 MHz symbol rate, both yielding 2 Mbps data rates. The endoscopy video obtained with a medical-grade system was transmitted with 16-QAM modulated signal with 1 MHz symbol rate, yielding 4 mbps data rates. In each case, a phase-coherent, decision feedback equalizer was used at the receiver end to equalize for the channel effects. The media transmitted and received throughout the experiment can be found in the \href{https://bit.ly/3wuc7tk}{referenced repository}\footnote{\url{https://bit.ly/3wuc7tk}}.

In this work, the transmit system was only responsible of interfacing with the camera and transmitting the signal with minimal processing. Most of the data processing, on the other hand, occurs at the receiver end. The receiver end is envisioned to be implemented as a software on a computer in a medical examination room, similar to an ultrasound machine. Therefore, the receiving end is more flexible in terms of processing power and footprint. However, more engineering work is needed to integrate the demonstrated transmission system with existing, small-footprint systems that could interface with a camera and transmit a QAM communication signal.

Currently, the latency of the video-streaming system is on the order of a few seconds. $40 \mu s$ of this latency is imposed by the noncausal part of the feedforward filter in the DFE. Reducing the filter size to eliminate this latency could cause performance degradation. However, the remaining part of the latency due to the FFMPEG, waveform generator, and digitizer memory buffers. Since the National Instruments equipment is controlled by a non-real time operating system, waveform generator and a video player only start after the memory buffers are filled up to a threshold in order to ensure streaming without a buffer underflow. The buffering latency can be reduced by deploying the system on a real-time processing environment.

A benchtop computer is used in this work to build a proof-of-concept system quickly and with better control. Although the footprint of the system is currently too large for an implantable device, the ultrasonic transducers used for transmission are suitable for use in a capsule-sized device. In addition to the transducer, a small camera, a controller board and a battery are envisioned to be part of the biomedical implant. All these additional parts exist separately as commercial devices in small form factors. When combined with such devices with further engineering work, the proof-of-concept system demonstrated in this paper could enable CE devices to have real-time video broadcasting and could be employed in active, online medical implants.

\ifCLASSOPTIONcaptionsoff
  \newpage
\fi
\balance
\bibliographystyle{IEEEtran}
\bibliography{meatcomms2_preprint}

\end{document}